\documentstyle[seceq, epsf]{ptptex}

\newcommand{\EQ}{\begin{equation}}
\newcommand{\EN}{\end{equation}}
\newcommand{\EQA}{\begin{eqnarray}}
\newcommand{\EQN}{\end{eqnarray}}
\newcommand{\EQAN}{\begin{eqnarray*}}
\newcommand{\EQNN}{\end{eqnarray*}}
\newcommand{\e}{{\rm e}}

\newcommand{\Tr}{{\rm Tr}}

\notypesetlogo  
\preprintnumber[3cm]{
UT-Komaba/03-7}

\title{
What is Holography  in the Plane-Wave Limit \\of the 
AdS$_5$/SYM$_4$ Correspondence ?
}

\author{
Tamiaki {\sc Yoneya}\footnote{E-mail: tam@hep1.c.u-tokyo.ac.jp. 
This expository article is based on talk 
``Holography and the Plane-Wave Limit of the 
AdS/CFT Correspondence" given by the 
present author at the Yukawa Memorial  Symposium, November, Nishinomiya, 2002. 
The present version is considerably expanded and 
corrected from the 
one to be published in the Proceedings. 
}
}

\inst{
Institute of Physics, University of Tokyo, Komaba 153-8902, Tokyo
}

\abst{
The issue of holographic principle in the 
PP-wave limit of the AdS/CFT correspondence 
is discussed, 
in the hope of clarifying some confusions 
in the literature. We show that, in the plane-wave limit, 
the relation between the partition function in the bulk and 
the gauge-invariant correlation functions  on the boundary 
should be interpreted on the basis of a tunneling picture 
in the semi-classical approximation 
which is appropriate for the plane-wave limit. 
This leads to a natural relation between 
{\it Euclidean} S-matrix in the bulk and the 
short-distance operator-product 
expansion 
of the so-called BMN operators on the boundary. 
}

\begin{document}

\maketitle

\section{Introduction}
About a year ago, a remarkable conjecture,  
extending the AdS/CFT 
correspondence\cite{maldacena} such that 
a class of non-BPS stringy states are included in the 
correspondence in a particular 
limit of plane-wave geometries,   
has been put forward by Berenstein, Maldacena, and 
Nastase (BMN)\cite{bmn}. 
In spite of many interesting works done 
along this conjecture,  however, 
the  formulation of  holographic principle\cite{thooft} in the 
plane-wave limit remains 
 still quite elusive.  It is important 
to understand this problem, 
since the holographic principle is believed to govern 
such correspondences between bulk quantum 
gravity and (gauge) field theories on the boundary. 
We have discussed this issue in our previous 
work\cite{dsy} 
appeared in {\it hep-th} archive half a year ago. 
In the present article, we would like to 
revisit this issue, since it seems  unfortunately 
that our previous work 
has not been appreciated sufficiently by most workers 
in this field.  I hope to make further clarification 
of our main arguments with some corrections to a part of 
discussions in the original version of our work.

We  point out some crucial puzzles, which 
I believe are lying in the heart of the problem, 
associated with 
the original proposal on the 
correspondence of a class of gauge-invariant 
local operators (BMN operators) on the Yang-Mills side 
with the string states defined along a null trajectory 
in the bulk.  Then  a simple tunneling 
picture is introduced, solving all of the 
puzzles. This  leads us to a proposal of 
a natural holographic 
correspondence, namely, the 
direct relation between the operator-product 
expansion (OPE) of the BMN operators 
defined on the boundary of the AdS 
geometry and the Euclidean infinite-time 
transition amplitudes (`Euclidean S-matrix') 
defined along a tunneling null geodesic 
connecting two-points on the boundary. 

\section{Basic holographic relation and the BMN 
limit}

Let us first briefly recall the basic relation\cite{holography} 
between the partition function in the bulk and 
the generating functional for correlation 
functions on the boundary, which has been 
conjectured to be valid in the supergravity 
limit of the AdS/CFT correspondence, 
\EQ
Z[\phi_0]_{{\rm gravity}}=\langle \exp (\int d^4 x 
\sum_i \phi^i_0(x){\cal O}_i(x) )\rangle_{{\rm ym}}. 
\label{relcorr}
\EN
The left-hand side is the partition function 
of supergravity with boundary 
conditions on the independent set of fluctuating 
fields $\{\phi_i\}$ in the bulk, 
\EQ
\lim_{z\rightarrow 0} \phi^i(z, x)= z^{4-\Delta_i}
\phi^i_0(x) ,
\label{boundc}
\EN
where the variable $z$ is defined to be vanishing 
at the boundary, using the Poincar\'{e} coordinates 
of the AdS$_5$ geometry, 
\EQ
ds_{P}^2 ={R^2dz^2 \over z^2} +{ d\vec{x}_3^2-dt^2\over R^2 z^2}.
\EN
 The right-hand side of (\ref{relcorr}) 
is the generating functional 
for an appropriate set of gauge-invariant 
operators $\{{\cal O}_i\}$ with definite conformal dimensions 
$\Delta_i$, which couple 
to the set of fields 
$\{\phi^i_0(x)\}$ at the boundary. The set  $\{{\cal O}_i\}$ 
must have a one-to-one correspondence 
to the set  $\{\phi^i_0\}$.  

Although we do not have any rigorous derivation 
of the above relation (\ref{relcorr}), 
there is a natural physical picture justifying it\footnote
{For further discussions on this picture, we invite the 
reader to the reference\cite{yorev}
}: 
This relation can be interpreted as two apparently 
different but equivalent descriptions of the 
low-energy behavior of a source-probe system of 
D3-branes. Suppose that probe D3-branes are put 
somewhere around the conformal boundary 
of the AdS$_5$ geometry which describes  the 
near horizon region of the background space-time 
of a large number ($N$) of source D3-branes. 
From the viewpoint of closed-string theory or 
supergravity as its low-energy approximation, 
the information of the probe D3-branes can be 
encoded into the boundary condition 
of the fluctuating fields in the bulk.  On the 
other hand, from the viewpoint of open-string 
theory or effective super Yang-Mills theory 
as its low-energy approximation, the information 
of the probe D3-branes can be encoded 
into the set of external fields coupled to appropriate 
gauge-invariant operators which corresponds 
to closed-string vertex operators for 
the bulk fields $\{\phi_i\}$.  The Yang-Mills description 
is justified if we restrict ourselves in the regime 
where the approximation  in which 
only the lowest modes of open strings 
attached to D3-branes are explicitly taken 
into account is effective. This  is natural at least  
in the extreme near horizon limit where 
the typical length scale of open strings are 
assumed much shorter than the string scale $\ell_s$. 
On the other hand, the distance scale of 
the AdS geometry is $R\equiv (4\pi g_sN)^{1/4}$ in the string 
unit $\ell_s=1$. Hence, to justify the low-energy 
approximation in the bulk in the weak string-coupling 
region $g_s\ll 1$, we have to take the large $N$ limit 
  such that 
$
R\gg 1. 
$
One point, which does not seem so natural in this conjecture 
but is actually of crucial importance, is 
that the description in terms of the lowest open 
string states must be justified in the whole near-horizon 
region where the length scale of open strings 
is small compared with the length scale $R\ell_s$ of 
curvature of the target space-time. It is not clear whether 
the typical length scale of open strings really remain 
small compared to  
the string scale when $R$ is large. In the original 
notation of ref.\cite{maldacena}, a typical energy scale 
$U=r/\ell_s^2$ is fixed, but 
it is tacitly assumed that the description is 
valid for arbitrarily large $U$, after we take the 
near horizon approximation 
$1+ R^4\ell_s^4/r^4=1+R^4/(U^4\ell_s)^4\rightarrow R^4/(U\ell_s)^4$. We expect that 
the superconformal symmetry  and the large $N$ limit is responsible for valid justification. 
To keep  this point in our mind, 
 we always use 
the string unit $\ell_s=1$  without 
taking the zero-slope limit.

In the supergravity limit, we are usually interested in 
the Kaluza-Klein modes of supergravity fields 
with respect to $S^5$, since the energies 
of stringy excited states of closed strings 
are much higher than them and 
hence are decoupled. If the magnitude of angular momentum 
along $S^5$ is $J$, the typical energy of KK modes is 
$J/R$, while the typical energy  of the string excitations 
is of order 1 in the string unit. Therefore, for finite $J$ 
the string excitations can indeed be 
ignored in the limit of large $R$. 
However, if we consider sufficiently large $J$, 
the string excitation energies can become comparatively 
smaller than those of KK modes, and 
we are not allowed to neglect them even if 
$R$ is taken to be large. Because of a large momentum 
associated with large $J$, we are no more in the 
naive low-energy regime. 
Since the masses of KK modes are related to the 
conformal dimensions, we expect that stringy excitation 
energies contribute to anomalous dimensions 
of generic (non BPS) string states. 
For example, for chiral supergravity modes, 
the conformal dimension  is equal to $J$, $\Delta =J$. 
Naively, the excitation energies $|n| \, \, (\sim 
O(1))$ contribute to the 
mass as $M^2 \sim (J/R)^2 + |n|$. A possible identification 
of conformal dimension would then be 
$
\Delta \sim M^2/(J/R^2) \sim J + R^2 |n|/J. 
$
This suggests that the anomalous dimension 
can be finitely detected in the large $R$ regime 
if one takes a double limit $R, J \rightarrow \infty$ 
by keeping the ratio $R^2/J$ finite.  

Remarkably, BMN showed that these qualitative considerations can be elevated to a much more 
precise theory. Combining this with 
the earlier observation due to 
Metsaev\cite{met} that 
the world-sheet formulation of string theory in the so-called PP-wave limit which just corresponds to 
the above double limit 
is exactly soluble 
as a free massive 2-dimensional field theory, 
they argued that the light-cone energy 
of the excitations of the form $\sqrt{1+R^4 n^2/J^2}$ 
is reproduced to all orders with respect to small 
$R^4/J^2$ expansion, using the large $N$ perturbation 
theory in the planar limit, if one 
identifies the gauge invariant operators corresponding to 
string oscillations 
appropriately.  Note that the above naive expectation 
for the form of $\Delta$ is the first order approximation 
with respect to large $R^2/J$ expansion corresponding to 
a flat-space limit which is opposite to large $N$ perturbation 
theory.  
In their interpretation, 
the longitudinal light-cone momentum is nothing but 
the rescaled angular momentum $P^+\sim J/R^2$ 
and the light-cone Hamiltonian is identified with 
$P^- \sim \Delta -J$. 
Schematically, the proposed correspondence 
is as follows: First choose two-directions, say, 
5 and 6, along $S^5$ in the  6 dimensional space 
which is orthogonal to D3-branes 
and is represented by 6 scalar fields on the Yang-Mills side. 
The angular momentum $J$ is associated with 
the rotation in the 5-6 plane. The chiral 
supergravity state, the ground state in the light cone, 
 is then identified with 
the local operator $\Tr(Z^J)$,  with 
$Z=(\phi_5+i \phi_6)/\sqrt{2}$, which indeed 
has the protected conformal dimension $\Delta=J$.  The bosonic stringy excitation 
modes are assumed to correspond to operators with 
various insertions of other 4 scalar fields  $\phi_i\, \, (i=1, \ldots, 4)$ 
and of 4 derivatives $D_iZ \, \, (i=1, \ldots, 4) $, each 
being accompanied by phase factor $\exp i (2\pi n \ell/J)$ 
where $\ell$ is the position of insertion and 
$n$ is the world-sheet momentum along the 
(closed) string. Thus the excited states with $n=0$ correspond 
to non-chiral (with respect to $J$) supergravity modes and 
8 physical transverse directions of string oscillations 
correspond to fields $\{\phi_i, D_i Z : i=1, \ldots, 4\}$. 
The fermionic modes are treated in a similar manner.   

Perhaps, one of curious features of this proposal 
would be that all of the  string excitations 
which are of course extended in 
bulk space-time correspond to {\it local} operators 
on the boundary, unlike naively expected Wilson-loop 
like operators. But this is not so surprising if one 
recalls that, on the Yang-Mills side, the 
bulk space-time actually corresponds to the 
configuration space of world-volume fields, and hence the 
extendedness in the bulk enters through the fluctuations of 
fields themselves, not as the extendedness 
with respect to the base-space coordinates of D3-branes. 
This phenomenon can also be regarded as a manifestation 
of the stringy uncertainty principle\cite{yo} of space-time 
as applied to D3-branes or its macroscopic 
version, the UV/IR relation\cite{susswitt}. 
It would be very interesting to investigate 
the whole picture of holography and 
its plane-wave limit 
using entirely the language of open strings 
before going to the effective SYM description.  
Such an approach might lead to a more 
direct and systematic derivation of 
the basic holographic relation on the basis 
of open-closed string duality. I hope to report 
progress along this line in a forthcoming work. 

\section{Puzzles and resolution}

Let us now reconsider the BMN proposal from the 
viewpoint of the basic holographic relation (\ref{relcorr}). 
If we first restrict ourselves to  supergravity approximation, 
a supergravity state with large angular momentum can 
be approximated by a semi-classical particle picture. 
Consider therefore the simplest scalar field equation 
in the AdS geometry using the WKB approximation; 
\EQ
\Big(z^2 \partial_z^2 -3 z\partial_z + R^4z^2 \omega^2
-J(J +4)\Big)\phi(z)=0 ,
\label{swaveeq}
\EN
where we have factorized the dependencies on the 
angular momentum $J$ and the (Minkowski) energy $\omega$. 
Assuming the standard WKB form 
$
\phi(z) \sim {\cal N}A(z)\exp iS(z) ,
$
we find the WKB phase $S(z)$ satisfies 
\EQ
z^2\Big({dS\over dz}\Big)^2-R^4z^2\omega ^2 
+J^2=0 .
\label{wkb1}
\EN
If we consider usual propagating solutions  
with real $S$, we have the condition 
\EQ
z^2\ge J^2/(\omega^2R^4). 
\EN
This inequality shows that for any finite energy $\omega$ and nonzero $J$, 
the particle trajectories (null geodesics) with finite ratio 
$J/R^2$ can never reach the boundary $z\rightarrow 0$. 
In the usual discussion of null geodesics in AdS space-times, 
it is convenient to use the global coordinates. 
The above property then corresponds to the fact that 
the null trajectory traversing a great circle along $S^5$ 
never reaches the conformal boundary and goes 
inside the horizon of the Poincar\'{e} patch in a finite interval with respect to 
the global time coordinate. 

This is puzzling since according to (\ref{relcorr}) 
the identification of bulk states with the 
Yang-Mills operators are made using boundary 
conditions near the conformal boundary $z=0$ of the 
AdS geometry. In fact, the behavior $z^{4-\Delta}$ 
in (\ref{boundc}) comes from the choice of 
the non-normalizable wave 
function in the classically forbidden region. 
In other words, the correspondence between the bulk and 
the boundary actually takes place as a  tunneling phenomenon 
in the semi-classical picture.  Indeed, by assuming 
purely imaginary action function 
$S\rightarrow iS_E, \, \, \phi(r) \rightarrow 
NA(z)\exp -S_E(z)$, we can easily check that 
there exist tunneling trajectories connecting 
two points on the boundary as solutions for 
Hamilton-Jacobi equation 
\EQ
z^2\Big({dS_E\over dz}\Big)^2= 
J^2(1-{z^2\omega ^2R^4 \over J^2}) \quad
\rightarrow \quad
S_E(z) \sim \pm J \log z, \quad A(z) \sim z^{2\mp 2} ,
\label{wkbsol}
\EN
as $z\rightarrow 0$ giving 
$
\Delta =J+4 \quad \mbox{or} \quad  -J. 
$
This gives the well known mass-dimension relation 
$m^2=J(J+4)=\Delta(\Delta -4)$ for scalar field. 
The former positive solution $\Delta =J+4$  
represents the {\it non}normalizable wave function 
which is responsible for the relation 
(\ref{relcorr}) and (\ref{boundc}).\footnote{
Note that {\it non}normalizability is with respect to 
the `inside' region corresponding to large $z$. If we take the viewpoint 
from the probe (`outside' region), namely, from $z=0$, 
the nonnormalizable solutions actually 
correspond to normalizable solutions. 
}
The explicit form of the tunneling trajectory 
is 
\EQ
z={J \over R^2\omega \cosh \tau} , 
\label{tunnelgeodesic}
\EN
which is obtained by replacing $dS_E/dz$ by the $z$-momentum 
$P_z= Jz^{-2}dz/d\tau$ and similarly by
 $J=P_{\psi}=Jd\psi/d\tau, \omega=-P_t=
J(R^4z^2)^{-1}dt/d\tau$, with $\tau$ being the 
affine parameter along the trajectory. 
See Fig. 1. 

\begin{center}
\begin{figure}
\begin{picture}(190,300)
\put(120,0){\epsfxsize 170pt  \epsfbox{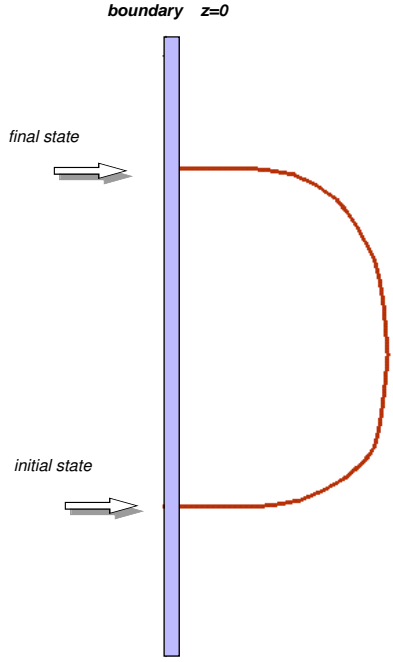}}
 \end{picture}
\caption{The tunneling null trajectory from 
boundary to boundary.}
\label{fig1} 
\end{figure}
\end{center}

Now we emphasize that our tunneling picture 
solves other puzzles related to the 
holographic interpretation. 
First, the 8 transverse directions 
in the BMN proposal includes the 4 base-space 
directions $(\vec{x}_3, t)$ of D3-branes. If we use the 
ordinary real null geodesics, 
the direction of the trajectory 
in the  region (corresponding to the turning point $z=J/\omega R^2$) where it becomes closest to 
the boundary goes parallel to the base space. 
This seems  contradictory to the identification 
of transverse directions along the trajectory of 
strings. On the contrary, in the tunneling picture, the trajectory becomes manifestly orthogonal to 
the base space directions of D3-branes, 
as we approach to the boundary. 
The above tunneling trajectory connects 
two points on the boundary. 
The coordinate distance between the two points at the 
boundary is asymptotically equal to $|r|=2{J\over \omega}$. 
Also, the natural affine time $\tau$ along the 
tunneling trajectory is related to 
the coordinate $z$,  $z \rightarrow {2J\over R^2\omega}\e^{-|\tau|} $ which 
gives a direct justification 
for the identification of 
energy with the conformal dimensions (minus $J$), 
in conformity with the conformal transformation 
law of the Poincar\'{e} coordinates. 

Furthermore, the tunneling picture gives 
a natural rationale for another subtlety 
involved in the original proposal. 
To have the real null geodesics, it is absolutely 
essential to use Minkowskian signature for the 
background space-time. But then the 4
 dimensional world volumes of D3-branes have also 
Minkowskian signature. We thus 
encounter again a contradiction to the identification of base-space directions with the 
transverse directions of light-cone formulation 
of strings. In our tunneling picture, however, the use of 
purely imaginary affine time, comparing 
with the case of propagating region, necessarily demands us to 
Wick-rotate the target time and angular 
coordinate simultaneously in order to 
keep $J$ and $\omega$ real in the 
WKB equation describing the tunneling 
region. Thus the 4 base space directions 
must be treated as Euclidean, while the angular 
variable $\psi$ along $S^5$ now replaces the role of 
target time coordinate. Incidentally, this 
 `double' Wick rotation of target space-coordinates 
required by the tunneling picture fits nicely with 
a similar formal prescription, as adopted in 
\cite{bs}, which is technically 
demanded in the boundary state 
formalism for D-branes in the light-cone gauge.

Our simple consideration on the nature 
of the semi-classical picture behind the 
holographic relation (\ref{relcorr}) 
clearly suggests that we should be able to 
extend the relation to include stringy 
states by replacing the null trajectory 
in the usual derivation of the plane-wave 
limit of string theory by the tunneling null 
trajectories which directly connect 
points on the conformal 
boundary.  
This picture originates from the general fact that, 
for nonzero $J$, probing D3-brane from
 the conformal boundary 
is inevitably a tunneling phenomenon, 
though our proposal does not exclude 
other possible approaches to holography.\footnote{A  representative work which is very different 
from out viewpoint is \cite{berenastase}. 
For a (necessarily, partial) list of other references,    
we refer the reader to the bibliography of \cite{dsy}. 
}   

In connection with this, our viewpoint 
also seems to 
resolve the question of introducing physical observables 
in terms of bulk string theory. If we insist that 
the theory is defined around the ordinary real 
null geodesic, the background trajectory goes 
inside  the horizon of D3-branes metric in a {\it finite}  light-cone 
time $\tau$. Remember that the
 real null trajectory is 
$z=J/(\omega R^2\cos \tau$) 
 in terms of the Poincar\'{e} 
coordinates. Then, it seems difficult to 
associate  scattering events occurring 
in the bulk 
with physical observables 
of the Yang-Mills theory in any 
systematic way, since we do not know 
how to use the global coordinates of 
AdS geometry in defining string theory 
observables associated with D-branes. 
The situation is very 
different according to our 
proposal: Boundary-to-boundary 
transitions involve infinite 
time duration with respect to 
natural affine (Wick-rotated in the above sense) 
time $x^+ \propto 
\tau$, in the limit that the boundary,  
on which we deal with the Yang-Mills 
theory observables coupled to 
external sources corresponding to 
probe branes, approaches to the conformal boundary of 
AdS geometry.  This is natural since 
only known manner of extracting physical 
observables in string theory is through scattering 
amplitudes corresponding to infinite time duration. 
In section 5, we propose a direct relation 
between such amplitudes, {\it Euclidean} S-matrix, 
 and the short-distance 
structure of  SYM correlators.

\section{String theory along a tunneling null geodesic}

Let us now briefly indicate how the string theory 
expanded around the tunneling null geodesic looks like. 
Since the details are given in our previous paper, 
we only present the final results correcting 
some errors in the original version 
of ref.\cite{dsy}.  
What we do is just the semi-classical expansion of the 
Green-Schwarz action around the tunneling 
trajectory (\ref{tunnelgeodesic}) with the Wick rotations 
discussed above. To simplify the expressions below, 
we choose the parameters of the trajectory 
such that $J=R^2 \omega$ and define the longitudinal 
momentum as $\alpha = J/R^2 (>0)$.  
 In the `standard' notation 
commonly used in the recent literature, our 
convention corresponds to 
set $\mu=1$. 
In the scaling limit, we can truncate the 
expansion to the second order after eliminating the 
negative metric associated with the Wick-rotated 
angle direction $\psi$ using the Virasoro constraint. The bosonic action 
\EQ
S_b = {R^2\over 2\pi}\int d\tau \int_0^{2\pi \alpha} d\sigma 
\, 
{1\over 2}\Big[
z^{-2}(\partial z)^2 + 
z^{-2}(\partial \vec{x}_4)^2 
-\cos^2 \theta (\partial \psi)^2
+(\partial \theta)^2 +
\sin^2\theta (\partial \Omega_3)^2
\Big] 
\EN
then reduces to that of free massive theory 
as 
\EQ
S^{(2)}_b =
{1\over 4\pi}\int d\tau \int_0^{2\pi \alpha}
d\sigma
\Big[
(\partial \vec{x}_4)^2 + \vec{x}_4^2 + (\partial \vec{y}_4)^2 + \vec{y}_4^2
\Big], 
\label{finalbosonicaction}
\EN
performing field redefinition and rescalings appropriately 
in the large $R$ limit, 
where $\vec{x}_4$ is the {\it redefined} 4-vector originating  
from the  fluctuations along the base space directions 
of D3-branes with suitable mixing with the $z$-direction 
inside the bulk ($z\ne 0$), and $\vec{y}_4$ is the 4-vector corresponding 
to the fluctuations along $S^5$ in the directions to 
orthogonal to the trajectory. This is the same action 
as we obtain in the case of real trajectory, 
except for the difference that 2-dimensional metric is 
Euclidean ($\partial^2 =\partial_{\tau}^2 + \partial_{\sigma}^2$). 

Similarly, the final form of the 
fermionic action is found to be 
\EQ
S_f^{(2)}=
{i\over 2\pi}\int d\tau \int_0^{2\pi \alpha}
d\sigma \, 
\Big[
\theta^I\Gamma_0\Gamma_-\partial_{\tau}
\theta^I
-is^{IJ}\theta^I\Gamma_0\Gamma_- \partial_{\sigma}
\theta^J
-i\epsilon^{IJ}\theta^I\Gamma_0
\Gamma_-\Pi
\theta^J\Big]
,
\label{fermionaction}
\EN
with 
\EQ
\Pi = i\Gamma_{0123}, \quad \Pi^2 =1 , \quad 
[\Pi, \Gamma_{\pm}]=0=\{\Pi, \Gamma_0\}, \quad \Pi^T=-\Pi, \quad \Pi^{\dagger}=\Pi, 
\EN
and 
with $I, J =1, 2$ and 
$
s^{IJ} =\pmatrix{ 1 & 0 \cr 
                                   0 & -1 \cr}. 
$
The fermionic coordinates satisfy 
\EQ
\Gamma_+ \theta^I =0,  \quad \Gamma_+\equiv
(\Gamma_Z -i \Gamma_{\psi})/\sqrt{2}=\Gamma_-^{\dagger} , 
\EN
in addition to the Weyl condition of type IIB theory. 
Because of double Wick rotation, the fermionic coordinates 
$\theta^I $ 
become complex with the same number of 
independent degrees of 
freedom (16=8+8) as the usual Majorana-Weyl 
spinors in the ordinary Minskowskian 10 dimensional 
space-time.  

The manifest global symmetry of the fermionic action 
is $SO(4)\times SO(4)$, due to the 
presence of the mass term containing the Gamma-matrix 
factor $\Pi$, while the bosonic action has manifest 
$SO(8)$ symmetry. However, we can trivially eliminate $\Pi$ 
by making the redefinition
\EQ
\theta^1 \rightarrow \theta^1, \theta^2 
\rightarrow \Pi\theta^2, 
\label{thetaredef}
\EN
since the kinetic term is 
invariant under this transformation. 
This makes manifest that the fermionic action has also the $SO(8)$ symmetry. Formally this redefinition 
is the same  as the one associated with T-duality 
transformation in flat space-time along the 
four base-space directions, since in the 
flat limit $\theta^{1,2}$ reduce to holomorphic and 
anti-holomorphic coordinates, respectively. Then, 
the enhancement of global symmetry seems natural 
because a D3-brane turns into a D-instanton. 
Though suggestive, this argument is   incomplete,  
since the bosonic part is not T-dualized.\footnote{
For a related discussion in flat (Minkowski) limit, see \cite{ima}. }
Naively, such an 
enhancement of symmetry may seem very strange,  
if we recall that the background RR-gauge field itself which 
is responsible to the fermionic mass term 
is obviously not invariant 
under the transformations mixing two different $SO(4)$ directions of AdS$_5$ and $S^5$, respectively. 
The situation is that in the plane-wave limit, 
the information on the direction of the RR-gauge 
field is lost and only its magnitude is detected by 
the strings at least in the particular 
case of the AdS$_5\times S^5$ geometry.\footnote{
As a simple analogy, consider a set of four $O(2)$ vector 
fields $(\phi_1, \phi_2), (\phi_3, \phi_4), 
(\psi_1, \psi_2), (\psi_3, \psi_4)$
and the potential 
$
V(\phi, \psi)=a\sum_{i=1}^4 (\phi_i^2 +\psi_i^2)
+b(\phi_1\psi_1+\phi_2\psi_2 -\phi_3\psi_3 -\phi_4\psi_4)
+c(\phi_1\psi_1+\phi_2\psi_2)^2 +
d(\phi_3\psi_3 +\phi_4\psi_4)^2 ,
$
which is symmetric under $O(2)\times O(2)$. 
If we take the limit of weak fields by neglecting 
the quartic terms,  the symmetry is 
trivially enhanced to $O(4)$ by reinterpreting 
$(\phi_1,\phi_2,\phi_3, \phi_4)$ 
and $(\psi_1, \psi_2, -\psi_3, -\psi_4)$ as two $O(4)$ vectors. 
If the quartic terms are not totally 
ignored, we can never have enhanced symmetry. }
In the following, we use the convention 
in which the factor $\Pi$ is eliminated 
for spinor coordinates and 8-vector notation 
$(\vec{x}_4, \vec{y}_4) \rightarrow 
(x^i; i=1, 2, \ldots, 8)$ for vector coordinates. 
It would be very interesting if we could see this 
enhancement of symmetry directly on the 
Yang-Mills side. Remember
 that the appearance of the $SO(8)$ is 
quite mysterious on the Yang-Mills side 
even if we restrict ourselves to purely bosonic excitations, 
in view of very different origin of 4+4 transverse 
modes in the BMN proposal. Note also that 
we can never see  the $SO(8)$ symmetry generators  
by simply contracting the symmetry algebra 
of the AdS$_5\times S^5$ geometry. 
We have to redefine the generators,  
correspondingly to the field redefinition (\ref{thetaredef}). 

To quantize the system, it is convenient to use 
manifestly $SO(8)$ conventions.  
Using the standard notation of $SO(8)$-$\gamma$ matrices ($8\times 8$), 
Hamiltonian, and CCR are given as 
\EQ
H={1\over 2}\int_0^{2\pi\alpha}
d\sigma 
:\Big[
2\pi p^2 +{1\over 2\pi}(x')^2 +{1\over 2\pi}x^2 
-{i\over 2\pi}(\theta\theta' +(2\pi)^2 
\lambda\lambda') +2\theta\lambda
\Big]: ,
\EN
\EQ
\theta=\theta^1+i\theta^2, \quad \lambda={1\over 2\pi}
\overline{\theta}={1\over 2\pi}(\theta^1-i\theta^2) ,
\EN
\EQ
[x^i(\sigma), p^j(\sigma')]=i\delta^{ij}\delta(\sigma-\sigma')
,\quad 
\{\theta_a(\sigma), \lambda_b(\sigma')\}=
\delta_{ab}\delta(\sigma-\sigma') .
\EN
General world-sheet fields $O(\tau, \sigma) $ are defined by 
$O(\tau, \sigma)=\e^{H\tau}O(\sigma)\e^{-H\tau}$ 
and hence adjoint operation  
is $O(\tau, \sigma)
\rightarrow O(-\tau, \sigma)^{\dagger}$. 
In this sense, the Euclidean theory 
satisfies physical positivity condition 
(or `reflection' positivity).

The Hamiltonian are 
diagonalized by the following mode expansions, suppressing vector and spinor indices,
\EQ
x(\sigma)=x_0+
\sqrt{2}\sum_{n=1}^{\infty}
(
x_n\cos {n\sigma\over \alpha}+
x_{-n}\sin {n\sigma\over \alpha}
), 
\EN
\EQ
p(\sigma)={1\over 2\pi \alpha}\Big[p_0+
\sqrt{2}\sum_{n=1}^{\infty}
(
p_n\cos {n\sigma\over \alpha}+
p_{-n}\sin {n\sigma\over \alpha}
)
\Big], 
\EN
\EQ
\theta(\sigma)=\theta_0+\sqrt{2}\sum_{n=1}^{\infty}
(
\theta_n\cos {n\sigma\over \alpha}+
\theta_{-n}\sin {n\sigma\over \alpha}
), 
\EN
\EQ
\lambda(\sigma)={1\over 2\pi \alpha}\Big[\lambda_0+
\sqrt{2}\sum_{n=1}^{\infty}
(
\lambda_n\cos {n\sigma\over \alpha}+
\lambda_{-n}\sin {n\sigma\over \alpha}
)
\Big], 
\EN
\EQ
x_n={i\over \sqrt{2\alpha E_n}}
(a_n-a_n^{\dagger}), \quad 
p_n=\sqrt{{\alpha E_n \over 2}}(a_n+a_n^{\dagger}), 
\EN
\EQ
\theta_n={1\over 2\sqrt{\alpha E_n}}
(\ell_n^+b_{n} -i \ell_n^- b_{-n}^{\dagger}), \quad
\lambda_n={1\over 2}\sqrt{{\alpha\over E_n}}
(i\ell_n^-b_{-n}+\ell_n^+b_n^{\dagger}), 
\EN
where 
\EQ
\ell_n^{\pm}=\sqrt{E_n+{n\over \alpha}}\pm 
\sqrt{E_n-{n\over \alpha}}, \quad 
E_n =\sqrt{1 +{n^2\over \alpha^2}} ,
\EN 
\EQ
[a_n, a_m^{\dagger}]=\delta_{nm} , 
\quad \{b_n, b_m^{\dagger}\}=\delta_{nm} ,
\EN
 for all (positive, negative and zero) integers $n$. 
We adopted the sine-cosine basis instead of the 
exponential basis of ref. \cite{dsy}
The Hamiltonian, to be identified with 
$P^-=E-J$ is simply 
$H=\sum_{n=-\infty}^{\infty}E_n(a_n^{\dagger}a_n 
+b_n^{\dagger}b_n)$. 
With the vacuum being defined by $a_n|0\rangle =
0=b_n|0\rangle$,  the $SO(8)$ symmetry is 
completely manifest. \footnote{
We note that our $SO(8)$-vacuum is slightly different from 
the one discussed in ref.\cite{mettsey} in the Minkowski 
formalism, because of our redefinition of the 
spinor coordinates.} 
In a forthcoming work, we hope to present an 
explicit construction of 
 string-field theory using our manifest $SO(8)$ formalism.  

The supersymmetry in this formalism is rather subtle. 
The following fermionic generators, 
which we propose to call `pseudo' susy generators, 
commute with the Hamiltonian, 
\EQ
R_{\dot{a}}^-= \int_0^{2\pi\alpha} d\sigma
\Big[(p+{i\over 2\pi}x)\cdot \gamma_{\dot{a}b}
\theta_{b}+{1\over 2\pi}x'\cdot \gamma_{\dot{a}b}
\overline{\theta}_b
\Big], \quad 
\overline{R_{\dot{a}}^-}=\Big(R_{\dot{a}}^-\Big)^{\dagger} , 
\EN
and satisfy the algebra, in the Hilbert space of 
translation invariant states with respect to 
$\sigma \rightarrow \sigma + \mbox{const.}$, 
\EQ
\{R_{\dot{a}}^-, \overline{R_{\dot{b}}^-}\}
=2\delta_{\dot{a}\dot{b}}H
-i\gamma^{ij}_{\dot{a}\dot{b}}L^{ij}
, \quad \{R_{\dot{a}}^-, R_{\dot{b}}^-\}=0 ,
\EN
with 
\EQ
L^{ij}=\int_{0}^{2\pi\alpha}d\sigma
\Big[
x^ip^j-x^jp^i +
{i\over 4\pi}\theta\gamma^{ij}
\overline{\theta}
\Big] . 
\label{L}
\EN
This algebra respects of course the SO(8) symmetry. 
However, the would-be SO(8) generator $L_{ij}$ has a 
wrong sign for the fermionic contribution. This means that 
the algebra actually does not close with a finite 
number of generators. Namely, we are necessarily lead to 
an infinite dimensional algebra, if we require that 
the algebra is covariant under SO(8). In particular, 
we have  to define an infinite number of fermionic generators to close the algebra. 

The standard susy generator which is 
covariant only under $SO(4)\times SO(4)$, 
can be conveniently expressed 
by making a canonical transformation to the 
spinor coordinates as 
\EQ
\psi={1\over 2}(1+\Pi)\theta + {1\over 2}(1-\Pi\overline){\theta}, \quad 
\overline{\psi}={1\over 2}(1-\Pi)\theta + 
{1\over 2}(1+\Pi)\overline{\theta}\EN
where,  in the standard $(8\times8)$ 
SO(8) spinor notation for gamma matrices 
($(\gamma_i\gamma_j^T+
\gamma_j\gamma_i^T)_{ab}=2\delta_{ab}, 
\, \, (\gamma_i^T\gamma_j+
\gamma_j^T\gamma_i)_{\dot{a}\dot{b}}=2
\delta_{\dot{a}\dot{b}}$) 
\EQ
\Pi_{ab}=\Pi_{ba}=(\gamma_1
\gamma_2^T\gamma_3\gamma_4^T)_{ab}, \quad 
\Pi_{\dot{a}\dot{b}}=\Pi_{\dot{b}\dot{a}}=(\gamma_1^T
\gamma_2\gamma_3^T\gamma_4)_{\dot{a}\dot{b}}. 
\EN
\EQ
Q_{\dot{a}}^-= \int_0^{2\pi\alpha} d\sigma 
\Big(
(p\cdot\gamma-{i\over 2\pi}x\cdot\gamma\Pi)\psi
-{1\over 2\pi}x'\cdot \gamma \overline{\psi}
\Big)_{\dot{a}} ,
\label{qminus}
\EN 
\EQ
\overline{Q}_{\dot{a}}^-= \int_0^{2\pi\alpha} d\sigma 
\Big(
(p\cdot\gamma+{i\over 2\pi}x\cdot\gamma\Pi)  \overline{\psi}
-{1\over 2\pi}x'\cdot \gamma \psi
\Big)_{\dot{a}} , 
\label{qminusbar}
\EN 
in terms of the spinor coordinates $\psi, \overline{\psi}$. 
Nontrivial part of the supersymmetry algebra is 
\EQ
\{Q_{\dot{a}}^-, \overline{Q}_{\dot{b}}^-\}= 2H\delta_{\dot{a}\dot{b}}
+\sum_{(i, j)\in(1,2,3,4)}i
(\gamma_{ij}\Pi)_{\dot{a}\dot{b}}J_{ij}
-\sum_{(i, j)\in(5,6,7,8)}i
(\gamma_{ij}\Pi)_{\dot{a}\dot{b}}J_{ij} ,
\label{dynamicalalge}
\EN
\EQ
J_{ij}=\int_0^{2\pi\alpha}d\sigma
\Big(
x_i p_j -x_jp_i -{1\over 4\pi}i\psi 
\gamma_{ij}\overline{\psi}
\Big) .
\EN
If we combine these 
standard susy generators with our SO(8), 
the algebra is again extended to an infinite 
dimensional algebra, corresponding to 
the above phenomenon. This suggests that we can have 
much stronger constraints on the dynamics of the 
system by combining  
 SO(8) and susy than taking into account only the 
standard supersymmetry and its smaller 
global symmetry SO(4)$\times$SO(4).  
It is certainly reasonable to expect that 
the dynamics respects  the SO(8) symmetry,  
which is hidden in the standard 
susy algebra but is exhibited in the world-sheet action. 
In any case, it seems very important to further 
clarify the role of the hidden SO(8) symmetry.

\section{Euclidean S-matrix and OPE}

We have obtained the above theory 
as a limit from the string theory on the AdS 
space-time such that the background 
trajectory of semi-classical expansion 
connects directly from conformal boundary to 
conformal boundary. Therefore, we are bound 
to have a natural extension of the 
basic holographic relation (\ref{relcorr}). 
An obvious guess for the left-hand side (namely 
bulk partition function with boundary 
conditions for physical fluctuating modes) 
of (\ref{relcorr}) for general string states 
is the S-matrix 
in the Euclidean sense describing infinite 
propagation of states from $\tau=-T$ to $\tau=T$ 
along the trajectory 
in the limit $T\rightarrow \infty$. 
What is the natural correspondent for the 
right-hand side?  In terms of the string language, 
putting boundary condition 
amounts to preparing initial and final multi-string 
states appropriately near  the ends of the 
background trajectory. 
On the Yang-Mills side, they must be 
associated with some products of BMN operators 
in an appropriate basis.   
Since the background trajectory meets 
the boundary only at two points, we expect that 
they are some sort of short distance products 
at each meeting points, initial and final points. 
However, it is not clear {\it a priori} how to define such 
short-distance products, fitting for the present 
expectation.\footnote{If we wish to 
consider more general $n$-point correlators, 
the semi-classical expansion around any 
single background trajectory is not sufficient: 
We have to necessarily  combine  $n$ different 
tunneling null geodesics with vertices 
describing branching of such trajectories. 
But that almost amounts to dealing with 
string field theory in the 
AdS background without relying on 
the plane-wave limit. Such general vertices 
cannot be described in light-cone gauge since 
the light-cone times are in general different for 
branching trajectories. }

Let us therefore first examine the structure of the 
Euclidean S-matrix in perturbation theory. 
Since the two-point S-matrix describing just the 
propagation of single string states can always be 
normalized to be identity, it is natural to define 
the perturbative expansion of the Euclidean S-matrix 
in analogy with the usual Minkowski case as 
\[\hspace{-5.8cm}
\langle b| (S-1)|a \rangle
=\lim_{T\rightarrow \infty}
\Big[-\langle b|{\cal V}|a\rangle
\int_{-T}^{T}d\tau 
\e^{(E_b-E_a)\tau}
\]
\[
+\sum_c \langle b | {\cal V}|c\rangle \langle c|{\cal V}
|a \rangle 
\int_{-T}^{T}d\tau 
\e^{(E_b-E_c)\tau}
\int_{-T}^{\tau} d\tau_1\e^{(E_c-E_a)\tau_1}
\]
\[
-\sum_c\sum_d 
\langle b | {\cal V}|c\rangle \langle c|{\cal V}
|d \rangle \langle d|{\cal V}
|a \rangle 
\int_{-T}^{T}d\tau 
\e^{(E_b-E_c)\tau}
\int_{-T}^{\tau} d\tau_1\e^{(E_c-E_d)\tau_1}\]
\EQ\hspace{5cm}\times 
\int_{-T}^{\tau_1} d\tau_2\e^{(E_d-E_a)\tau_2}
 + \cdots
\Big] ,
\label{tmatrix}
\EN
where ${\cal V}$ is the interaction part of the 
Hamiltonian of string field theory.

Take the simplest nontrivial case of 
3-point matrix elements, 
corresponding to $1\rightarrow 2$ or $2\rightarrow 1$ 
scattering in the tree approximation. 
Then only the first term contributes 
\EQ
 { \e^{(E_b-E_a)T}- \e^{-(E_b-E_a)T}\over
E_a-E_b }\, \langle b| {\cal H}^{(1)}|a\rangle 
\label{firstterm}
\EN
with 
${\cal V}\rightarrow {\cal H}^{(1)}$ being 
3-point interaction vertex of string fields. 
The singularity exhibited in this expression 
takes the following form, because of the conservation of 
angular momentum $J_k-J_i-J_j=0$, 
\EQ
{V_{ijk}\over (\Delta_k-\Delta_i -\Delta_j)}
\e^{\pm (\Delta_i +\Delta_j-\Delta_k)T}, 
\EN
where $(i, j)$ and $k$ correspond to 
2 and 1 string states, respectively, and 
the $V_{ijk}$ is the matrix element of 
the 3-point vertex. We can derive this same 
form on the basis of the standard
 LSZ formalism using Euclidean Green functions.  
Only difference from the Minkowski case is that 
the factor in front of the on-shell 
matrix element $\langle b| {\cal H}^{(1)}|a\rangle$ replaces the $\delta$-function of energy conservation. 
One might wonder why the nonconservation 
of energy is allowed in time-translation invarint theory. 
However, for scattering events, 
invariance under time translation comes only after 
integration over interaction times. 
In Minkowski case, the integration over the 
interaction times yields the energy-conserving 
$\delta$-function. But in the Euclidean case, 
that gives the singularity exhibited by this factor. 
 
Now, the above form should be compared with the 
standard form of the OPE, 
\EQ
O_i(0)O_j(x)\sim \sum_k {1\over |x|^{\Delta_i+\Delta_j-\Delta_k}}C_{ijk}O_k(0), \quad 
x\rightarrow 0.
\label{ope}
\EN 
Remembering that two-point 
amplitudes are normalized to be one, 
this strongly suggests us to identify the 
short distance cutoff 
as 
\EQ
|x| =e^{-T}
\label{cutoff}
\EN
We suppose that the standard OPE form 
(\ref{ope})  is valid on the 
Yang-Mills side for an appropriate 
basis of the BMN operators. In general, we should expect 
certain operator mixing of the original BMN operators. 
 Then provided that we can choose only 
the term with appropriate sign on the exponential 
(\ref{firstterm}) in the prefactor, we can identify the bulk S-matrix and correlation functions by 
setting 
\EQ
V_{ijk}=(\Delta_k-\Delta_i-\Delta_j)C_{ijk}. 
\label{vcrelation}
\EN
The relation (\ref{cutoff}) is natural from what 
we have discussed in section 2 with respect to 
the UV/IR relation. See Fig. 2. 
\begin{center}
\begin{figure}
\begin{picture}(200,190)
\put(110,0){\epsfxsize 230pt  \epsfbox{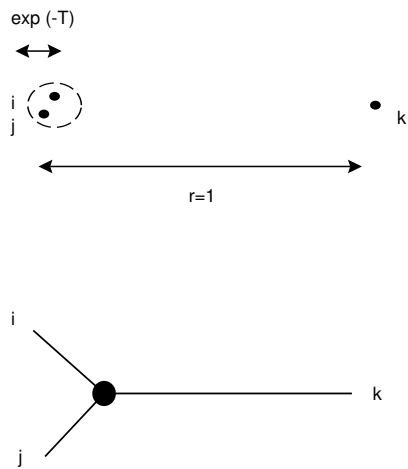}}
 \end{picture}
\caption{The short distance product and 3-point vertex. }
\label{fig1} 
\end{figure}
\end{center}

Let us now consider the meaning of the special choice 
of the sign for the exponentials in the singularities appearing in 
(\ref{firstterm}). In the case of ordinary S-matrix 
in Minkowski space-times, 
such a  choice of signs on the exponentials 
would amount to the familiar $i\epsilon$ prescription. 
In our Euclidean case, we have no such formal 
arguments. The above choice of sign 
on the exponent amounts to 
assuming  that the initial or final states 
{\it corresponding to two strings} have always 
positive exponent $\e^{ET}$ with 
$E$ being the total energy $E=E_i + E_j$. Here we recall that 
the perturbative 
expansion (\ref{tmatrix}) of S-matrix 
comes from the formal definition 
\EQ
\lim_{T\rightarrow \infty}
\langle b| \e^{{\cal H}^{(0)}T}\e^{-2{\cal H}T}
\e^{{\cal H}^{(0)}T}|a \rangle
\label{sdef}
\EN
where  ${\cal H}$ and ${\cal H}^{(0)}$ 
are full and free string field Hamiltonian. 
The factors $\e^{{\cal H}^{(0)}T}$ sandwiching 
transition operator $\e^{-2{\cal H}T}$ 
correspond to amputating the propagators 
of external lines in the language of LSZ formalism. 
The choice of the positive exponent for two-string 
states means that,   
the initial or final states, $\e^{{\cal H}^{(0)}T}|a\rangle$ or $\langle b|\e^{{\cal H}^{(0)}T}$,  with two strings must be 
prepared such that we observe 
the interactions of two strings into a single 
string (or reversed one) 
occurring near the conformal boundary associated with
the  initial (or final) 
region of the trajectory, respectively. 
Thus we arrive at the following 
correspondence between the processes 
in string theory and in Yang-Mills theory, respectively;
\[
\mbox{[interaction of multi-string states near the boundary]}
\] \[
\Updownarrow\]
\[
\mbox{[short distance product of multiple BMN operators]}  
\]
Preparation of the initial and final multi-string 
states as required above is possible 
by using wave packet basis along the 
trajectory appropriately. Note that 
the scattering we are dealing with  is essentially 
of 1+ 1 dimensional, since the wave functions 
are bounded in all the transverse directions  by 
harmonic potentials. Therefore the wave packet picture 
can be formulated in an elementary way, with a caveat
 that we are actually using it 
 in the sense of Euclidean theory 
with the steepest descent approximation 
for momentum ($J$) integrations in performing superposition 
of plane-`wave' functions. We expect that the tunneling 
picture would be more naturally formulated if the 
near horizon limit is not assumed, but we avoid such 
technical refinements in the present exposition. 

In terms of an obvious symbolic notation ${\cal C}$ for the 
string-field theoretic expression 
for the coefficients $C_{ijk}$, 
we can express the equation (\ref{vcrelation}) symbolically 
as 
\EQ
{\cal H}^{(1)} =\pm[{\cal H}^{(0)}, {\cal C}]. 
\label{h1comm}
\EN 
for $2\rightarrow 1$  or $1\rightarrow 2$ 
matrix elements, respectively. 
Here and in what follows, 
the products of symbolic string-field  operators 
denoted by calligraphic letters 
are to be interpreted within the restriction of 
semiclassical tree approximation 
of string field theory. We have assumed that 
the CFT coefficients are consistent with 
the kinematical symmetries such as our $SO(8)$ 
or, if one wishes, smaller $SO(4)$, 
$[{\cal J}, {\cal C}]=0$. 
In terms of 
the similar symbolic notation, we can easily see that 
the dynamical susy anticommutation relation is 
preserved by choosing the first order interaction 
correction to the dynamical susy generator as 
\EQ
{\cal Q}^{-(1)}=\pm [{\cal Q}^{-(0)}, {\cal C}] 
\EN
for the same matrix elements $2\rightarrow 1$  or $1\rightarrow 2$ as for those of the interaction Hamiltonian. 

In fact, we can reverse the above arguments, starting from 
suspersymmetry: 
 Assume a general form of  the first order interaction 
Hamiltonian as a square of a component ${\cal Q}^{(0)}$ 
of the lowest order susy generator, 
\EQ
{\cal H}^{(0)}=\Big({\cal Q}^{(0)}\Big)^2 .
\label{square}
\EN
In our case, $Q^{(0)}$ can be any component of 
the second-quantized version of 
$Q_{1}^-=(Q^-+\overline{Q}^-)/\sqrt{2}
$ or of $Q_{2}^-=(Q^--\overline{Q}^-)/\sqrt{2}i$. 
Using only the relation (\ref{square}), we can derive   
\EQ
[{\cal Q}^{(0)}, {\cal H}^{(1)}]
=[{\cal H}^{(0)}, {\cal Q}^{(1)}], 
\quad 
{\cal H}^{(1)}=\{{\cal Q}^{(0)}, 
{\cal Q}^{(1)}\}  . 
\EN
For any matrix elements which do not 
conserve energy ($E_i+E_j-E_k\ne 0$), this is 
rewritten as 
\EQ
\langle k|{\cal Q}^{(1)}|i, j\rangle 
=\langle k|[{\cal Q}^{(0)},{\cal C}]|i, j\rangle, 
\EN
with
\EQ
\langle \ell|{\cal C} |i, j\rangle 
={\langle \ell|{\cal H}^{(1)}|i,j\rangle , 
\over E_{\ell}-E_i -E_j } 
\label{h2comm}
\EN
for arbitrary $2\rightarrow 1$ matrix elements,  
since ${\cal Q}^{(0)}$ has nonzero matrix 
elements only when energy is conserved, $[{\cal Q}^{(0)}, 
{\cal H}^{(0)}]=0$. 
This implies that the holographic relation 
(\ref{h1comm}) is 
inevitable, supposing that there is no nontrivial 
{\it energy-conserving}  
matrix elements for the 
interaction Hamiltonian, since then 
we can adopt (\ref{h2comm}) as the 
definition of ${\cal C}$.  
The last condition seems indeed to be satisfied at least 
in the supergravity approximation \cite{slee}, and also 
to be consistent with all the known perturbative 
results \cite{pert} on the gauge theory side in the 
small $R^2/J$ limit.\footnote{
One might wonder if this argument is applied to 
simpler systems, such as (first quantized) 
supersymmetric 
quantum mechanics. In the latter case, there are in general 
an infinite number of nonzero energy-conserving 
matrix elements 
for the interaction part. Hence, we would not have a well-defined 
candidate for the operator ${\cal C}$. 
}
Conversely, our discussion shows that holographic 
principle requires the vanishing of 
$\langle \ell|{\cal H}^{(1)}|i,j\rangle$, 
when energy is conserved. We also note that 
the above argument is valid if 
we replace the susy generators by our 
pseudo-susy generators. 

These  are  main messages derived from our 
considerations on how the basic 
holographic relation (\ref{relcorr}) of supergravity 
is extended to string theory in the plane-wave 
limit. 
Our result is directly based on our 
interpretation of holography and is 
completely independent of the 
arguments given in ref. \cite{constetal}  where the 
relation (\ref{vcrelation}) 
of the same form as ours were first 
discussed in a context which 
is nothing to do with our arguments.  
It should also be emphasized that our 
Euclidean prescription is deeply motivated by the 
tunneling picture and is not just 
a formal device for computations 
of Minkowskian amplitudes.  
For more detailed discussions on our results,  
including the limitation of our methods, 
we refer the reader to 
the paper \cite{dsy}. For example, by assuming that 
the BMN operators form a complete set of 
gauge-theory operators 
with respect to OPE (\ref{ope}), we 
can extend above consideration to a more 
general $1\rightarrow n+1 \, (n+1\rightarrow 1)$ 
matrix elements, and 
the result is 
\EQ
{\cal H}={1\over 1-{\cal C}}{\cal H}^{(0)}(1-{\cal C}), 
\quad 
{\cal Q}^{-}={1\over 1-{\cal C}}{\cal Q}^{-(0)}
(1-{\cal C}),
\EN
for $1\rightarrow n+1$ and  their transpose  
for $n+1\rightarrow 1$, respectively. 

\section{Concluding remarks}
To avoid 
confusion, 
we note that the simple structure we found above does not necessarily 
mean that the interaction can completely be eliminated 
by a unitary transformation in full fledged quantum 
theory of string fields. 
In particular, our arguments are not sufficient 
for fixing the matrix elements of more 
general types than those discussed above.  
It should, however,  be stressed that 
the above form clearly shows that the 
implementation of susy algebra alone in the sense of 
{\it classical} string field theory  
is not sufficient to fix the 
interaction vertices  uniquely, since 
we can always construct the 3-point interaction vertex formally 
such that it satisfies 
the first order form of the susy algebra, 
once a ${\cal C}$ preserving kinematical 
symmetries is given. Of course, 
for the result to be well behaved, 
the CFT coefficients should not diverge for the special 
cases when the energy is conserved. In other words, 
the 3-point vertex must vanish for such matrix 
elements. 
In the case of supergravity approximation, this 
is indeed satisfied before taking the 
plane-wave limit, as has been emphasized before. 
It seems that all the known results for CFT 
coefficients from perturbative 
computations\cite{pert} on the Yang-Mills side are 
also consistent with this property. 
Another remark is that the relation (\ref{vcrelation}) 
does not necessarily require that the so-called prefactor of the 
string-field theory 
is equal to the energy difference, contrary to 
some earlier works done using a wrong 
expression for a possible form of string field theory 
vertex. For references to these earlier works, we 
refer the reader to the bibliography of our 
paper\cite{dsy}. Finally, 
we would like to mention a more recent proposal\cite{chuk}, 
appeared after the Symposium,   
of a prefactor which is different from the 
standard one\cite{sp} which leads to 3-point vertex 
 consistent with 
our general prediction and with the known 
perturbative results on the SYM side. \footnote{
In preparing the present manuscript, I came to know 
a more recent and alternative discussion \cite{div} related to this issue. They maintain the relation (\ref{vcrelation}) 
without taking into account the mixing effect. 
It is unfortunate that many authors 
in this area do not seem aware of our work. 
The reference of the present exposition itself is 
of course  very incomplete, since the purpose of this article 
is not to give a review. 
I  would like to apologize any authors whose works are 
overlooked here. 
}
In any case, however, our consideration seems to indicate 
that the logic of light-cone string field theory 
{\it alone} does not involve the first principles 
in order to fix the interacting string field 
theory in general curved space-times. 
At the same time, it is also important to 
have an explicit construction of string field 
theory which meets our criterion as 
holographic string field theory. We hope to 
report progress along such direction 
in a forthcoming work.




\section*{Acknowledgements}
I would like to thank the organizers of the Yukawa Memorial 
Symposium for invitation and the city of Nishinomiya for 
supporting such wonderful academic meetings for 
many years,  and 
S. Dobashi and H. Shimada for discussions. 
The present work is supported in part by Grant-in-Aid for Scientific 
Research (No. 12440060 and No. 13135205)  from the Ministry of  Education,
Science and Culture. 

\appendix


\end{document}